\documentclass[prc,showpacs,showkeys,aps,epsfig]{revtex4}

\usepackage{psfig}

\def\lsim{\raise0.3ex\hbox{$<$\kern-0.75em\raise-1.1ex\hbox{$\sim$}}}

\def\gsim{\raise0.3ex\hbox{$>$\kern-0.75em\raise-1.1ex\hbox{$\sim$}}}

\newcommand{\be}{\begin{equation}}

\newcommand{\ee}{\end{equation}}

\def\beq{\begin{equation}}

\def\eeq{\end{equation}}

\def\beqa{\begin{eqnarray}}

\def\eeqa{\end{eqnarray}}

\newcommand{\rr}{\mbox{\boldmath $r$}}

\newcommand{\rb}{\mbox{\boldmath $b$}}

\def\gappeq{\mathrel{\rlap {\raise.5ex\hbox{$>$}}

{\lower.5ex\hbox{$\sim$}}}}

\def\lappeq{\mathrel{\rlap{\raise.5ex\hbox{$<$}}

{\lower.5ex\hbox{$\sim$}}}}

\def\Toprel#1\over#2{\mathrel{\mathop{#2}\limits^{#1}}}

\newcommand{\rk}{\mbox{\boldmath $k$}}

\begin{document}

\title{ Exclusive vector meson production in electron-ion collisions }
\author{ V.P. Gon\c{c}alves $^{1}$, M.S. Kugeratski $^{2}$, M.V.T. Machado$^{3}$
and  F.S. Navarra$^2$}
\affiliation{$^{1}$ Instituto de F\'{\i}sica e Matem\'atica,  Universidade
Federal de Pelotas\\
Caixa Postal 354, CEP 96010-090, Pelotas, RS, Brazil\\
$^2$Instituto de F\'{\i}sica, Universidade de S\~{a}o Paulo,
C.P. 66318,  05315-970 S\~{a}o Paulo, SP, Brazil\\
$^3$ Centro de Ci\^encias Exatas e Tecnol\'ogicas, Universidade Federal do Pampa \\
Campus de Bag\'e, Rua Carlos Barbosa. CEP 96400-970. Bag\'e, RS, Brazil\\} 

\begin{abstract}

We calculate the nuclear  cross section for  coherent and  incoherent vector meson 
production within the QCD color dipole picture, including saturation effects.  
Theoretical estimates for scattering on both  light and heavy nuclei are given 
over a wide range of energy.

\end{abstract}



\maketitle

\vspace{1cm}

\section{Introduction}

One of the main theoretical expectations for the high energy regime of  Quantum 
Chromodynamics (QCD) is the saturation of the parton densities in hadrons and nuclei 
at small values of the Bjorken-$x$ variable and the formation of a Color Glass 
Condensate (CGC). This is one of the main topics of hadron physics  to be explored  in the new accelerators, such as the LHC and  possibly the future 
electron-ion collider.  The search for signatures of the CGC has been  subject of an active 
research  (for recent reviews see, e.g. \cite{cgc}), with the experimental data being successfully described by phenomenological models based on saturation physics \cite{gllm,GBW,bgbk,kowtea,iim,fs}.  However, more definite 
conclusions are not yet possible. In order to discriminate between these 
different approaches and test the CGC physics, it would be very important to consider 
an alternative search. To this purpose, the future 
electron-ion colliders offer a promising opportunity 
\cite{raju_ea1,raju_ea2}.

In a series of papers \cite {kgn1,kgn2,ccgn1,ccgn2} we  investigated the prospects of  
observing the CGC in a future electron-ion collider (For related studies see 
Refs. \cite{raju_ea4,raju_ea5,nik_ea}). As it 
has been already emphasized in these papers, the advantage of using nuclear targets is that the 
saturation scale $Q_s$ is much larger and this is crucial for the observation of most 
of the CGC effects. After these studies our conclusion was that it is very difficult 
to disentangle CGC effects from the standard linear QCD looking only at  the nuclear 
inclusive observables, such as the nuclear structure functions $F_2^A$, $F^{c,\,A}_{2}$ 
and $F_{L}^A$.
On the other hand, the study of diffractive observables was shown to be promising, as  
demonstrated in Refs. \cite{kgn2,ccgn2}.  In particular, in \cite{ccgn2} we found out that 
the saturation models are able to describe the current data on the nuclear structure 
function $F_2^A$ and predict that the contribution of diffractive events to the total 
cross section should be of $\approx 20 \%$ at large $A$ and small $Q^2$. In the asymptotic 
limit of very high energies diffractive events are expected to form  half of the total cross 
section, with the other half being formed by all inelastic processes \cite{nik_dif}.  
This observation motivates a more detailed study of diffractive processes. In particular,  
we can expect that the study  of  exclusive processes, such as  diffractive vector meson 
production, can   be useful to determine the QCD dynamics at high energies. It is important 
to emphasize that the experimental HERA data on  vector meson production in $ep$ processes 
are successfully described by the phenomenological saturation models (See, e.g., \cite{mara,kowtea,sandapen,kmw,pesc,watt08}). In contrast, very little is known about vector meson 
production off nuclei at high energies. Exclusive vector meson production in $e\,A$ 
interactions can be classified as coherent or incoherent. If the reaction  leaves the 
target intact, the process is usually called coherent. Otherwise it is called incoherent.  
These two types of diffractive hadron production were studied recently in 
Refs. \cite{tuc1,tuc2,tuc3,tuc4,marquet} using the CGC formalism. The main conclusion  
of these works  is that  coherent production can serve as a sensitive probe of the high 
energy dynamics of  nuclear matter, while  incoherent production measures the fluctuations  
of the nuclear color fields. These findings  are an additional motivation for our  study 
of  nuclear vector meson production.
We  focus our analysis on   exclusive vector meson production by real and  virtual photons,  
which  may help us to understand several  physical issues, besides saturation effects. 
Among these we find, for example,  the transition from the soft dynamics 
(at low virtualities of the photon $Q^2$) to the hard perturbative regime at high $Q^2$ 
and  the relative weights of coherent and incoherent interactions between the projectile 
and the target  nucleus.

In this paper  we calculate the coherent and incoherent production cross sections of vector 
meson production  using the dipole approach and a nuclear saturation model based on CGC 
physics. The main input for our calculation  is the dipole-target cross section, 
$\sigma_{dip}^{\mathrm{target}} = \sigma_{dip}(x,\rr)$,  which is determined by the QCD 
dynamics at small $x$. In the eikonal approximation it is  given by:
\begin{equation} 
\sigma_{dip} (x, \rr) = 2 \int d^2 \rb \,  {\cal N} (x, \rr, \rb)
\label{sdip}
\end{equation}
where $ {\cal N} (x, \rr, \rb)$ is the forward dipole-target scattering amplitude for a dipole with 
size $\rr$ and impact parameter $\rb$ which encodes all the
information about the hadronic scattering, and thus about the
non-linear and quantum effects in the hadron wave function (see e.g. \cite{cgc}). It  can be obtained by 
solving the BK (JIMWLK) evolution equation in the rapidity $Y \equiv \ln (1/x)$. 
Many groups have studied the numerical solution of the BK equation, but 
several improvements are still necessary  before using the solution in the calculation 
of  observables. In particular, one needs to include the next-to-leading order 
corrections into the evolution equation and perform a global analysis of all 
small $x$ data. It is a program in progress (for recent results see \cite{alba,alba2}). 
In the meantime it is necessary to use phenomenological models for 
$ {\cal N}$ which capture the most essential properties of the solution. Following \cite{ccgn2} we will use in our calculations  the model proposed in Ref. \cite{armesto}, which describes  the current scarce experimental data on the nuclear structure function as well as includes the  impact parameter dependence in the dipole nucleus cross section. In this model the forward dipole-nucleus amplitude is given by
\begin{eqnarray}
{\cal{N}}^A(x,\rr,\rb) = 1 - \exp \left[-\frac{1}{2}  \, \sigma_{dp}(x,\rr^2) \,T_A(\rb)\right] \,\,,
\label{enenuc}
\end{eqnarray}
where $\sigma_{dp}$ is the dipole-proton cross section and $T_A(\rb)$ is the nuclear profile function, which is 
obtained from a 3-parameter Fermi distribution for the nuclear
density normalized to $A$ (for details see, e.g., Ref. \cite{vicmag_hq}).
The above equation, based on the Glauber-Gribov formalism \cite{gribov},  sums up all the multiple elastic rescattering diagrams of the $q \overline{q}$ pair
and is justified for large coherence length, where the transverse separation $r$ of partons in the multiparton Fock state of the photon becomes a conserved quantity, {\it i.e.} the size of the pair $r$ becomes eigenvalue
of the scattering matrix. It is important to emphasize that for very small values of $x$, other diagrams beyond the multiple Pomeron exchange considered here should contribute ({\it e.g.} Pomeron loops) and a more general approach for the high density (saturation) regime must be considered. However, we believe that this approach allows us to estimate the magnitude of the high density effects in the  kinematical range of the future $eA$ colliders.

In the present work, we shall use two models for the  dipole-proton cross section $\sigma_{dp}$. One is
the very popular GBW model \cite{GBW}, which
interpolates between the small and large dipole configurations,
providing color transparency behavior, $\sigma_{dp}\sim \rr^2$,
as $\rr \rightarrow 0$ and constant behavior, $\sigma_{dp}\sim
\sigma_0$, at large dipole separation. This model is no longer able to describe the most recent HERA data and it has 
been replaced by other parameterizations. However we shall keep using it as a baseline, 
which will be compared with other dipole-proton cross sections, giving us an estimate of the 
sensitivity of the observable to changes in the dipole cross sections. 
The parameterization of the dipole-proton cross section is given by  the eikonal-like 
form \cite{GBW},

\begin{eqnarray}
\sigma_{dp}^{\mathrm{GBW}} (\tilde{x}, \,\rr^2)  =  \sigma_0 \,
\left[\, 1- \exp \left(-\frac{\,Q_{\mathrm{s}}^2(\tilde{x})\,\rr^2}{4} \right) \, 
\right]\,, \hspace{1cm} Q_{\mathrm{s}}^2(\tilde{x})  =  \left( \frac{x_0}{\tilde{x}}
\right)^{\lambda} \,\,\mathrm{GeV}^2\,,
\label{gbwdip}
\end{eqnarray}
where the saturation scale $Q_{\mathrm{s}}^2$ defines the onset of the
saturation phenomenon, which depends on energy, and $\tilde{x}= (Q^2 + 4\,m_f^2)/(Q^2 + W_{\gamma N}^2)$ (See \cite{GBW} for details). 
One of the 
drawbacks of the GBW model is that it has no impact parameter dependence. 
It is assumed that the impact parameter  dependence of $ {\cal N}$ can be 
factorized as ${\cal N}(x, \rr, \rb) = {\cal N}(x, \rr) S(\rb)$ and this last 
function is integrated over $\rb$, giving rise to the parameter $ \sigma_0$.

During  the last years an intense activity in the area resulted  in  more 
sophisticated dipole proton cross sections, which had more  theoretical 
constraints and which were able to give a better description of the more recent 
HERA and/or RHIC data \cite{kmw,pesc,watt08,kkt,dhj,gkmn,buw}. In what follows we will use the b-CGC 
model proposed in Ref. \cite{kmw}, which improves the IIM model 
 \cite{iim} with  the inclusion of  the impact parameter dependence in the dipole 
proton cross sections. The parameters of this model were recently fitted to describe 
the current HERA data \cite{watt08}.  Following \cite{kmw} we have that the dipole-proton cross section is given by:
\begin{equation}
\sigma_{dp}^{bCGC} (x,\rr^2) \equiv \int \, d^2 \bar{\rb} \, \frac{d 
\sigma_{dp}}{d^2  \bar{\rb}} 
\label{new_iim}
\end{equation}
where 
\begin{eqnarray}
\frac{d \sigma_{dp}}{d^2 \bar{\rb}} = 2\,\mathcal{N}^p(x,\rr,\bar{\rb}) =  2 \times 
\left\{ \begin{array}{ll} 
{\mathcal N}_0\, \left(\frac{ r \, Q_{s}}{2}\right)^{2\left(\gamma_s + 
\frac{\ln (2/r Q_{s})}{\kappa \,\lambda \,Y}\right)}  & \mbox{$r Q_{s} \le 2$} \\
 1 - \exp^{-a\,\ln^2\,(b \, r \, Q_{s})}   & \mbox{$r Q_{s}  > 2$} 
\end{array} \right.
\label{eq:bcgc}
\end{eqnarray}
with  $Y=\ln(1/x)$ and $\kappa = \chi''(\gamma_s)/\chi'(\gamma_s)$, where $\chi$ is the 
LO BFKL characteristic function.  The coefficients $a$ and $b$  
are determined uniquely from the condition that $\mathcal{N}^p(x,\rr)$  and its 
derivative 
with respect to $rQ_s$  are continuous at $rQ_s=2$. 
In this model, the proton saturation scale $Q_{s}$ now depends on the impact 
parameter:
\begin{equation} 
  Q_{s}\equiv Q_{s}(x,\bar{\rb})=\left(\frac{x_0}{x}\right)^{\frac{\lambda}{2}}\;
\left[\exp\left(-\frac{\bar{b}^2}{2B_{\rm CGC}}\right)\right]^{\frac{1}{2\gamma_s}}.
\label{newqs}
\end{equation}
The parameter $B_{\rm CGC}$  was  adjusted to give a good 
description of the $t$-dependence of exclusive $J/\psi$ photoproduction.  
Moreover the factors $\mathcal{N}_0$ and  $\gamma_s$  were  taken  to be free. In this 
way a very good description of  $F_2$ data was obtained. 
The parameter set  which is going to be used here is the one presented in the second 
line of Table II of \cite{watt08}:  $\gamma_s = 0.46$, $B_{CGC} = 7.5$ GeV$^{-2}$,
$\mathcal{N}_0 = 0.558$, $x_0 = 1.84 \times 10^{-6}$ and $\lambda = 0.119$.

The paper is organized as follows. In next section (Section \ref{sec:vm}) we discuss the coherent and incoherent vector meson production and present our main formulas. In Section \ref{sec:res} we present our predictions for the energy and $Q^2$ dependences  of the  $\rho$ and $J/\Psi$ total cross sections considering the coherent and incoherent cases for two different nuclei. Moreover, we estimate the ratio between the coherent and incoherent cross sections. Finally, in Section \ref{sec:conc} we summarize our main conclusions.

\section{Exclusive vector meson production in the color dipole approach}
\label{sec:vm}

In the color dipole approach the exclusive vector meson production 
($\gamma^* A \rightarrow VY$) in electron-nucleus interactions at high energies can be 
factorized in terms of the fluctuation of the virtual photon into a $q \bar{q}$ color 
dipole, the dipole-nucleus scattering by a color singlet exchange  and the recombination into 
the final state vector meson. This process is characterized by a rapidity gap in the final 
state. If the nucleus scatters elastically, $Y = A$, the process is called coherent 
production. On the other hand, if the nucleus scatters inelastically, i. e. breaks up ($Y = X$),   
the process is denoted incoherent production. In this paper we will  consider the 
color dipole description of the $\gamma^* A \rightarrow VY$ ($V=\rho,J/\Psi$) process, 
which is quite successful for the proton case \cite{mara,kowtea,sandapen,kmw,pesc,watt08,kop0} and can be extended 
to nuclei targets with the Glauber-Gribov formalism (For  recent reviews see \cite{review_nik,review_iva}). To a large extent we follow the 
pioneering papers 
\cite{Nikolaev:1992si,Benhar:df,Kopeliovich:1993gk,Kopeliovich:1993pw,Nemchik:1994fq,nem1,nem2}, 
where these issues were first addressed and also their further developments 
\cite{Nikolaev:1999bq,Benhar:ue,kop1,kop2,kop3}.  To 
be more precise, in this work we will update the calculations presented in 
\cite{gon_ma04}, extending them also to electroproduction and treating  more carefully the 
two production mechanisms, coherent and incoherent. For this we shall use  the formalism 
described in detail in Ref. \cite{kop1}. For the dipole cross section  
we shall use the parameterization introduced in \cite{watt08}, which was able to give a very 
good  description of data on exclusive vector meson production on proton targets. In this 
sense our work is an extension of  \cite{watt08} to nuclear targets.

As discussed above, the exclusive vector meson production in electron-nucleus interactions 
can be classified as coherent or incoherent.
If the reaction  leaves the target intact, the process is 
usually called coherent, and the mesons produced at different longitudinal 
coordinates and impact parameters add up coherently. The corresponding integrated cross 
section  is given in the high energy regime (large coherence length: $l_c \gg R_A$) by \cite{kop1}
\begin{eqnarray}
\sigma^{coh}\, (\gamma^* A \rightarrow VA) = \int d^2\rb \left\lbrace | \int d^2\rr
 \int dz \Psi_V^*(\rr,z) \, \mathcal{N}^A(x,\rr,\rb)\, \Psi_{\gamma^*}(\rr,z,Q^2)
|^2\right\rbrace
\label{totalcscoe}
\end{eqnarray}
On the other hand, the diffractive incoherent production of vector mesons off nuclei, 
$\gamma^* A \rightarrow VX$, is associated with the breakup of the nucleus. In this case 
one sums over all final states of the target nucleus, 
except those that contain particle production. The $t$ slope is the same as in the case of a 
nucleon target. Therefore we have: 
\begin{eqnarray}
\sigma^{inc}\, (\gamma^* A \rightarrow VX) = \frac{|{\cal I}m \, 
{\cal A}(s,\,t=0)|^2}{16\pi\,B_V} \;
\label{totalcsinc}
\end{eqnarray}
where at high energies ($l_c \gg R_A$) \cite{kop1}:
\begin{eqnarray}
|{\cal I}m \, {\cal A}(s,\,t=0)|^2  =  \int d^2\rb \, T_A(\rb) \left[ |\int d^2\rr \int dz 
\, \Psi_V^*(\rr,z) \, \sigma_{dp} \, \exp[- \frac{1}{2} \, \sigma_{dp} \, T_A(\rb)] \, 
\Psi_{\gamma^*}(\rr,z,Q^2)|^2 \right] 
\label{totalcsinc1}
\end{eqnarray} 
The $q\bar{q}$ pair attenuates with a constant absorption cross 
section, as in the Glauber model, except that the whole exponential is averaged 
rather than just the cross section in the exponent.  
The coherent and incoherent cross sections depend  differently on  $t$.  
As discussed in \cite{Kopeliovich:1993pw}, at small-$t$ ($-t\,R_A^2/3 \ll 1$) coherent 
production dominates the leptoproduction of vector mesons, with the signature being a sharp 
forward diffraction peak. On the other hand, incoherent production will dominate the vector 
meson production at large-$t$ ($-t\,R_A^2/3 \gg 1$), with the $t$-dependence being to a good 
accuracy the same as in the production off  free nucleons. Therefore, it is expected that 
this signature will allow us  to separate the two contributions even at a limited resolution 
in $t$. However, at an electron-ion collider,  if the nucleus remains intact in a small-$t$ 
interaction, it will escape too close to the beam to be detectable. This  implies that the 
experimental separation between  coherent and incoherent production will be a challenging 
task. In this paper we  focus our study on the energy and $Q^2$ dependence of  the total 
coherent and incoherent cross sections and their relative weights. We postpone the study of 
the $t$-dependence of these cross sections for a future publication.

In the Eqs. (\ref{totalcscoe}) and (\ref{totalcsinc1}) the functions 
$\Psi^{\gamma}_{h, \bar{h}}(z,\,\rr)$ and $\Psi^{V}_{h, \bar{h}}(z,\,\rr)$  
are the light-cone wavefunctions  of the photon and vector meson, respectively. 
The quark and antiquark helicities are labeled 
by $h$ and $\bar{h}$
  and reference to the meson and photon helicities are implicitly understood. The 
variable $\rr$ defines the relative transverse
separation of the pair (dipole) and $z$ $(1-z)$ is the
longitudinal momentum fraction of the quark (antiquark). 
In the dipole formalism, the light-cone
 wavefunctions $\Psi_{h,\bar{h}}(z,\,\rr)$ in the mixed
 representation $(r,z)$ are obtained through two dimensional Fourier
 transform of the momentum space light-cone wavefunctions
 $\Psi_{h,\bar{h}}(z,\,\rk)$ 
(see more details, e.g. in Ref. \cite{sandapen}). The
 normalized  light-cone wavefunctions of  longitudinally ($L$) and
 transversely ($T$) polarized photons are given by \cite{kmw}:
\begin{eqnarray}
\Psi^{L}_{h,\bar{h}}(z,\,\rr)& = & \sqrt{\frac{N_{c}}{4\pi}}\,
\delta_{h,-\bar{h}}\,e\,e_{f}\,2 z(1-z)\, Q \frac{K_{0}(\varepsilon r)}{2\pi}\,,
\label{wfL}\\
\Psi^{T(\gamma=\pm)}_{h,\bar{h}}(z,\,\rr) & = & \pm
\sqrt{\frac{N_{c}}{2\pi}} \,e\,e_{f}
 \left[i e^{ \pm i\theta_{r}} (z \delta_{h\pm,\bar{h}\mp} -
(1-z) \delta_{h\mp,\bar{h}\pm}) \partial_{r}
+  m_{f} \,\delta_{h\pm,\bar{h}\pm} \right]\frac{K_{0}(\varepsilon r)}{2\pi}\,,
\label{wfT}
\end{eqnarray}
where $\varepsilon^{2} = z(1-z)Q^{2} + m_{f}^{2}$. The quark mass
$m_f$ plays the  role of a regulator when the photoproduction
regime is reached.  Namely, it prevents non-zero argument for the
modified Bessel functions $K_{0,1}(\varepsilon r)$ towards $Q^2\rightarrow 0$.
 The electric charge of the quark of flavor $f$ is given by $e\,e_f$.

For vector mesons, the light-cone wavefunctions are not known
in a systematic way and have to be determined in a phenomenological way.  
Here, we follow the approach proposed in Refs. \cite{nem1,nem2} and discussed in detail in \cite{sandapen, dgkp:97, kmw}. It is assumed that the spin and polarization of the vector mesons  are the same as in the photon, which is considered as being predominantly a quark-antiquark state. Consequently, the wavefunctions for a transversely and longitudinally  polarized vector meson  are  given by
\begin{eqnarray}
  \Psi^V_{h\bar{h},\lambda=\pm 1}(r,z) =
  \pm\sqrt{\frac{N_c}{2\pi}}\, \frac{1}{z(1-z)} \, 
  \left\{
  \mathrm{i}e^{\pm \mathrm{i}\theta_r}[
    z\delta_{h,\pm}\delta_{\bar h,\mp} - 
    (1-z)\delta_{h,\mp}\delta_{\bar h,\pm}] \partial_r \, + \, 
  m_f \delta_{h,\pm}\delta_{\bar h,\pm}
  \right\}\, \phi_T(r,z).
  \label{tspinvm}
\end{eqnarray}
and 
\begin{equation}
  \Psi^V_{h\bar{h},\lambda=0}(r,z) = \sqrt{\frac{N_c}{2\pi}}\,
  \delta_{h,-\bar h} \,
  \left[ M_V\,+ \, \delta \, \frac{m_f^2 - \nabla_r^2}{M_Vz(1-z)}\,  
    \right] \, \phi_L(r,z),
  \label{lspinvm}
\end{equation}
where $\nabla_r^2 \equiv (1/r)\partial_r + \partial_r^2$ and $M_V$ is the meson mass.  As emphasized in \cite{sandapen}, the longitudinally polarized wave function is slightly more complicated due to the fact that the coupling of the quarks to the meson is non-local, contrary to the photon case.

\begin{table}
\begin{center}
\begin{tabular}{||lccccc||}
\hline
\hline
$V(m_V)$ & $m_f$ & $R^2$  & $N_L$  & $ N_T$ &  $\hat{e}_f$ \\
 (MeV) &  (GeV) & (GeV$^{-2}$) & &  &   \\
\hline
\hline
$\rho\,(776)$ &  0.14 & 12.9 & 0.853 & 0.911 &  $1/\sqrt{2}$ \\
$J/\Psi\,(3097)$ & 1.4 & 2.3 & 0.575 & 0.578 &  2/3 \\
\hline
\hline
\end{tabular}
\end{center}
\caption{Parameters and normalization
  of the boosted-Gaussian overlap function.}
\label{tab:1}
\end{table}

Following Ref. \cite{pesc} we define the overlap function between the photon and vector meson wave functions  by
\begin{eqnarray}
\Phi_{\lambda}^{\gamma^*V}(z,\rr ;Q^2,M_V^2)
 = \sum_{fh\bar{h}} \left[
   \Psi^{V,\lambda}_{f,h,\bar{h}}(z,\rr ;M_V^2)\right]^*
\Psi^{\gamma^*,\lambda}_{f,h,\bar{h}}(z,\rr ;Q^2)\ .
\label{eq:wf}
\end{eqnarray} 
Using the above expressions for the photon and vector meson wave functions we  get  (See appendix of Ref. \cite{pesc}):
\begin{eqnarray}
\Phi^{\gamma^*V}_L(z,\rr,Q^2)  
   & = & \hat{e}_f \sqrt{\frac{\alpha_e}{4\pi}} N_c \, 2Q K_0(\varepsilon r)  
\left[M_V z(1-z)\phi_L(\rr,z)+ \frac{m_f^2-\nabla_r^2}{M_V}\phi_L(\rr,z)\right],\\
\Phi^{\gamma^*V}_T(z,\rr,Q^2) 
   & = & \hat{e}_f \sqrt{\frac{\alpha_e}{4\pi}} N_c \frac{\alpha_e 
N_c}{2\pi^2}\left\{
             m_f^2 K_0(\varepsilon r)\phi_T (r,z)
           - [z^2+(1-z)^2] \varepsilon K_1(\varepsilon r)) \partial_r\phi_T(r,z)
            \right\},
\end{eqnarray}
where 
\begin{eqnarray}
\phi_{L,T} = N_{L,T}
\,\exp\left[-\frac{m_f^2R^2}{8z(1-z)}+\frac{m_f^2R^2}{2}-\frac{2z(1-z)r^2}{R^2}
\right]\ .
\end{eqnarray}
and the parameters $R$ and $N_{L,T}$ are constrained by unitarity of the
wavefunction as well as by the electronic decay widths. The  parameters  used in our calculations are  presented in  Table \ref{tab:1}. Our choice is usually denoted in literature as {\it boosted Gaussian} (BG) wavefunctions and are 
a simplified version ($\delta = 1$) of the wavefunctions proposed by Nemchik, Nikolaev, Predazzi and Zakharov   in Refs. \cite{nem1,nem2}.
  We quote Refs. \cite{sandapen,kmw,pesc}
for more details  and  comparison with data for  both photo and electroproduction of 
vector mesons in $ep$ collisions at HERA.

Using the above expressions for the vector meson wave functions, we obtain 
\begin{eqnarray}
|{\cal I}m \, {\cal A}_{L,T}|^2 & = & \int d^2\rb \, T_A(\rb) \left\lbrace |\int d^2\rr \int dz 
\, \Phi^{\gamma^*V}_{L,T}(z,\rr,Q^2)   \,\sigma_{dp} \exp[- \frac{1}{2}\sigma_{dp} \,  T_A(\rb)]|^2 
\right\rbrace \,\,,
\label{inc}
\end{eqnarray}
and
\begin{eqnarray}
\sigma^{coh}\, (\gamma^*_{L,T} A \rightarrow VA) & = &\int d^2\rb \left\lbrace | 
\int d^2\rr \int dz \, \Phi^{\gamma^*V}_{L,T}(z,\rr,Q^2)  \left[ 2 \left(1 - \exp[- \frac{1}{2}\sigma_{dp}\,  T_A(\rb)]\right)
\right]|^2\right\rbrace \,\,.
\label{coe}
\end{eqnarray}
In what follows we will estimate the total cross section for the exclusive vector meson 
production  using  as input in the Eqs.  (\ref{inc}) and (\ref{coe}) the GBW and bCGC 
parameterizations  for the dipole-proton cross section.

\begin{figure}
\vspace{1.0cm}
\centerline{\psfig{figure=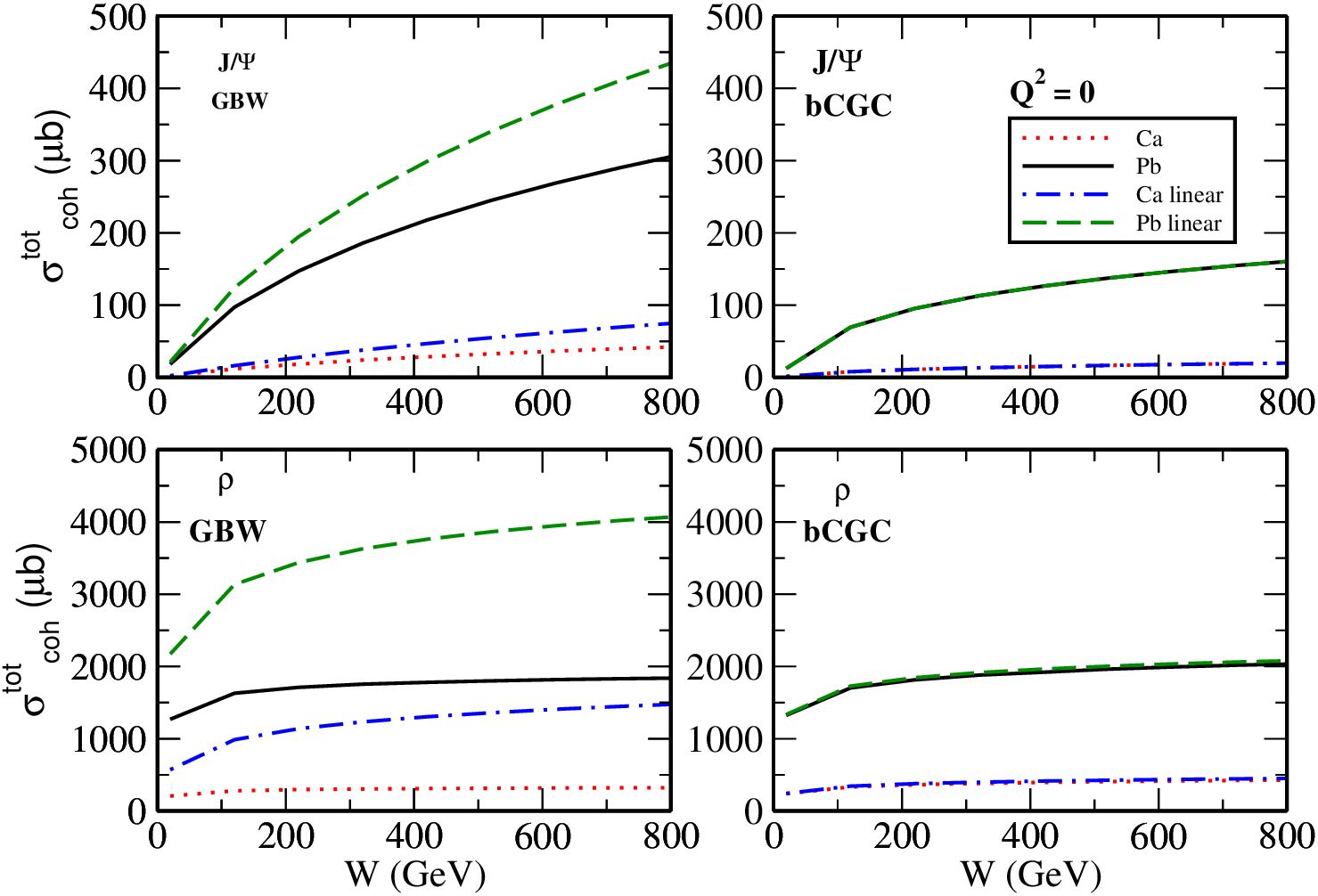,width=16cm}} 
\vspace{0.5cm}
\caption{Coherent photoproduction cross sections for $J/\psi$ and $\rho$ vector meson production
 with the GBW (left panels) and bCGC (right panels)  models. }
\label{fig1}
\end{figure}

\section{Results}
\label{sec:res}

\begin{figure}
\vspace{1.0cm}
\centerline{\psfig{figure=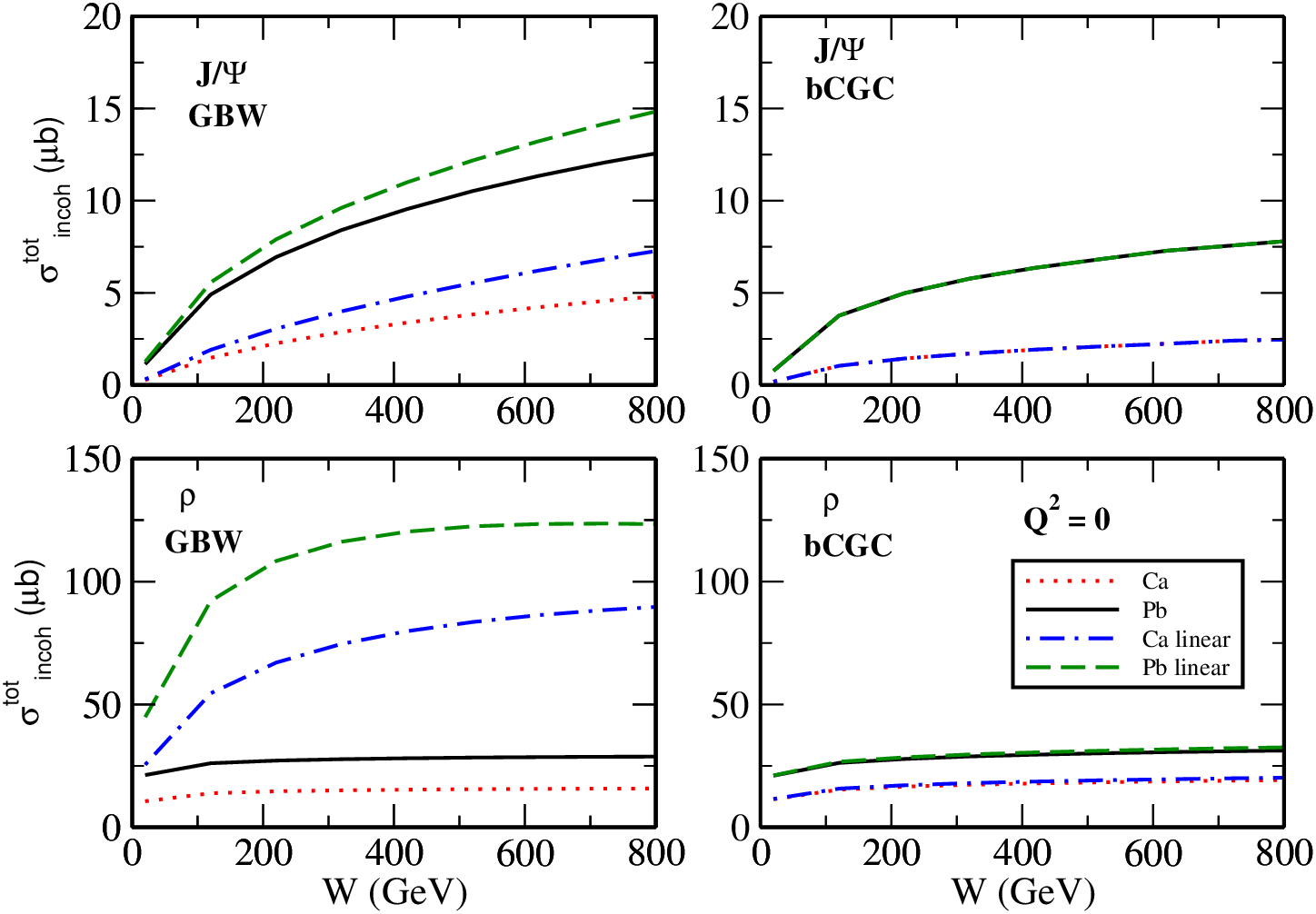,width=16cm}} 
\vspace{0.5cm}
\caption{Incoherent photoproduction cross sections for $J/\psi$ and $\rho$ vector meson 
production
 with the GBW (left panels) and bCGC (right panels)  models. }
\label{fig2}
\end{figure}

The project of an electron-ion collider is being elaborated by groups at BNL (eRHIC) 
\cite{raju_ea2} and at CERN (LHeC) \cite{dainton}. The main expectation is that the future 
$eA$ collider should have a center of mass energy larger than 60 $GeV$ and a high luminosity, 
such that measurements of  inclusive observables at low-$x$ and diffractive processes could be 
improved substantially. Here we will analyze the energy and $Q^2$ dependence of the coherent 
and incoherent total cross section for $J/\Psi$ and $\rho$ vector meson production. We will 
consider two different nuclei ($Ca$ and $Pb$ ) in order to  study the $A$-dependence of our 
predictions. 

Moreover, we will also use as input in our calculations the linear limit of the GBW and bCGC 
dipole-proton cross section in order to determine the sensitivity of our results to the 
saturation effects in the proton. In the GBW model we take the linear limit  by 
using instead of $\sigma_{dp}$, the expansion of 
(\ref{gbwdip}) when the dipole is very small ($r \rightarrow 0$) and the exponent is 
small. In this case $\sigma_{dp} =  \sigma_0 \, 
\frac{\,Q_s^2(\tilde{x})\,\rr^2}{4}$. In the bCGC model the linear behavior is estimated by 
taking only the first line of Eq. (\ref{eq:bcgc}).

\begin{figure}
\vspace{1.0cm}
\centerline{\psfig{figure=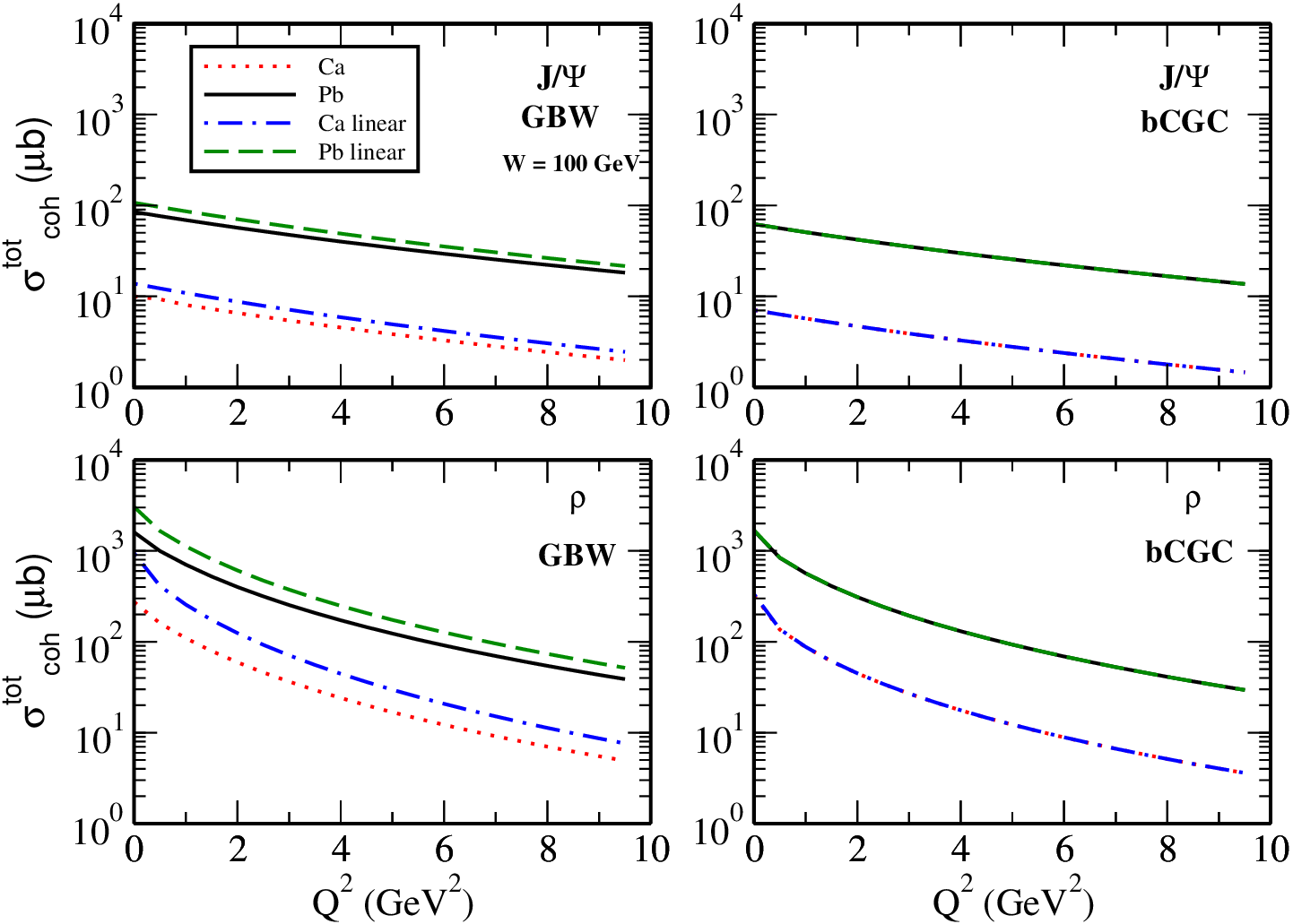,width=16cm}} 
\vspace{0.5cm}
\caption{Coherent electroproduction cross sections for $J/\psi$ and $\rho$
 with the GBW (left panels)  and bCGC (right panels)  models as a function of $Q^2$ at fixed 
center of mass energy ($W = 100$ GeV). }
\label{fig3}
\end{figure}

Let us start our analysis considering  nuclear photoproduction of vector mesons. 
In Figs. \ref{fig1} and \ref{fig2} we present our predictions for the energy dependence 
of the coherent and incoherent total cross sections, respectively.  As expected, the 
cross section grows with the energy and with the atomic number. We predict huge coherent 
cross sections, in reasonable agreement with those presented in \cite{gon_ma04}, where a 
different approach was used to estimate the cross sections.  
At fixed atomic number, the coherent cross section is approximately a 
factor 70 (25) larger than the incoherent one for $\rho$ ($J/\Psi$) photoproduction. 
Furthermore, at fixed energy, the coherent cross section increases with the atomic 
number faster than the incoherent one, in agreement with the conclusions from \cite{tuc4}.

The photoproduction of $\rho$ in nuclear collisions is a typical soft process, 
characterized by large dipole pair separations. At high energies, the saturation scale 
can assume large values and become the hard scale of the problem. The 
transition between these two regimes is determined by the energy dependence of the 
saturation scale, which is model dependent. From Figs. \ref{fig1} and \ref{fig2} we can see 
that for the $\rho$ meson the full predictions of the GBW and bCGC models are very similar, 
in contrast with the linear ones. While the cross sections are strongly dependent on   
assumptions concerning  saturation in the proton in the GBW case, they are very similar when 
we use the linear and full expressions of the bCGC model. This behavior is directly associated 
to the large difference between the linear and full expressions in the GBW model at large 
dipole pair separation, which is smaller in the bCGC model. Concerning $J/\Psi$ 
photoproduction, which is characterized by a hard scale, the mass of $J/\Psi$, the total 
cross sections increase faster with the energy than in  the case of the $\rho$.  Moreover, 
the GBW and bCGC predictions are very different, which is directly associated with the energy 
dependence of the saturation scale in these two models. Similarly to the $\rho$ case, the 
linear and full GBW predictions are different, while are similar when we use the bCGC model.

In Figs. \ref{fig3} and \ref{fig4} we present our predictions for the photon virtuality $Q^2$ 
dependence of the coherent and incoherent total cross sections, respectively. As expected, the 
total cross sections decrease with $Q^2$. The $Q^2$ behavior predicted by the GBW and bCGC 
models are similar. The main difference is associated to the strong dependence on the 
saturation effects in the proton  observed  when we use the GBW model. Moreover, 
the $Q^2$ dependence of the $J/\Psi$ cross section is smaller than the $\rho$ one, which is 
directly associated to the hard scale present in the $J/\Psi$ case.

\begin{figure}
\vspace{1.0cm}
\centerline{\psfig{figure=sectotincjpsirho_w100models.eps,width=16cm}} 
\vspace{0.5cm}
\caption{Incoherent electroproduction cross sections for $J/\psi$ and $\rho$
 with the GBW (left panels) and bCGC (right panels)  models as a function of $Q^2$ at fixed center of mass energy ($W = 100$ GeV). }
\label{fig4}
\end{figure}

\begin{figure}
\vspace{1.0cm}
\centerline{\psfig{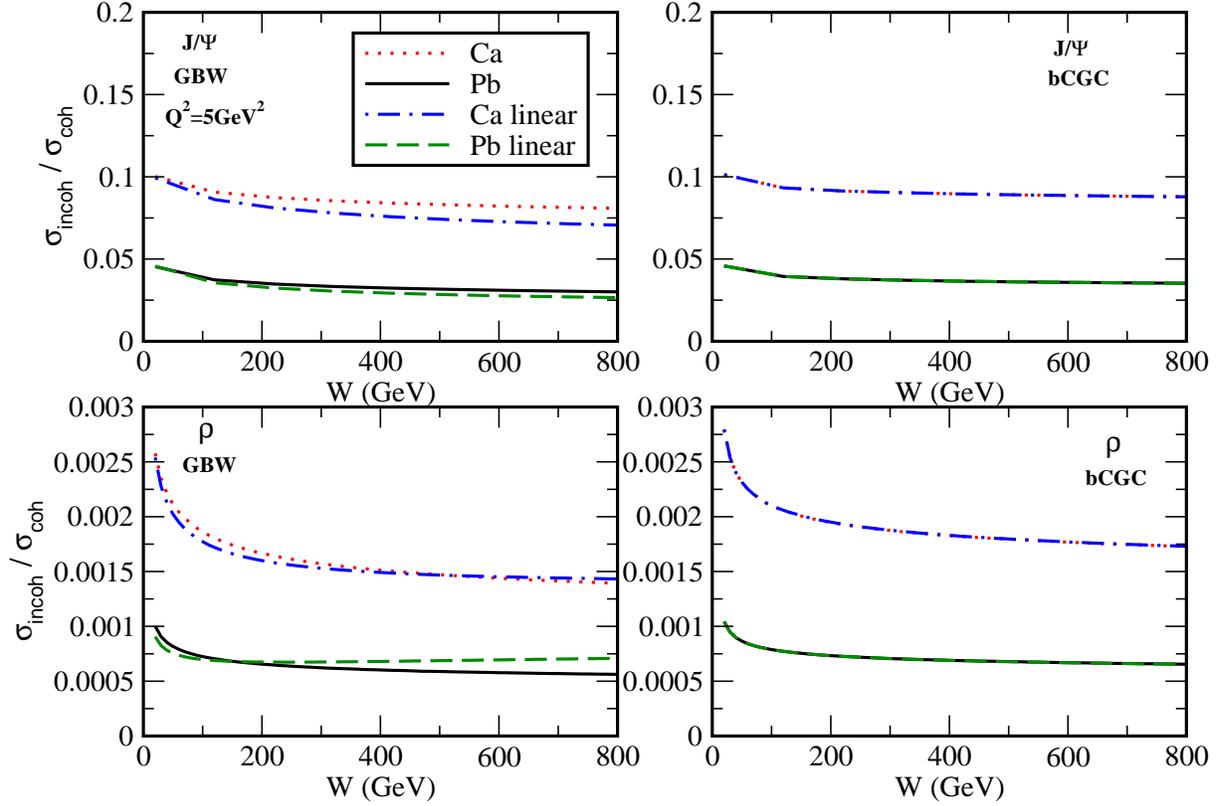}} 
\vspace{0.5cm}
\caption{Ratio $\sigma_{incoherent}/ \sigma_{coherent}$ for the nuclear electroproduction of 
vector mesons as a function of the energy with the GBW (left panels)  and bCGC (right panels) 
models at fixed virtuality ($Q^2 = 5$ GeV$^2$).}
\label{fig5}
\end{figure}

\begin{figure}
\vspace{1.0cm}
\centerline{\psfig{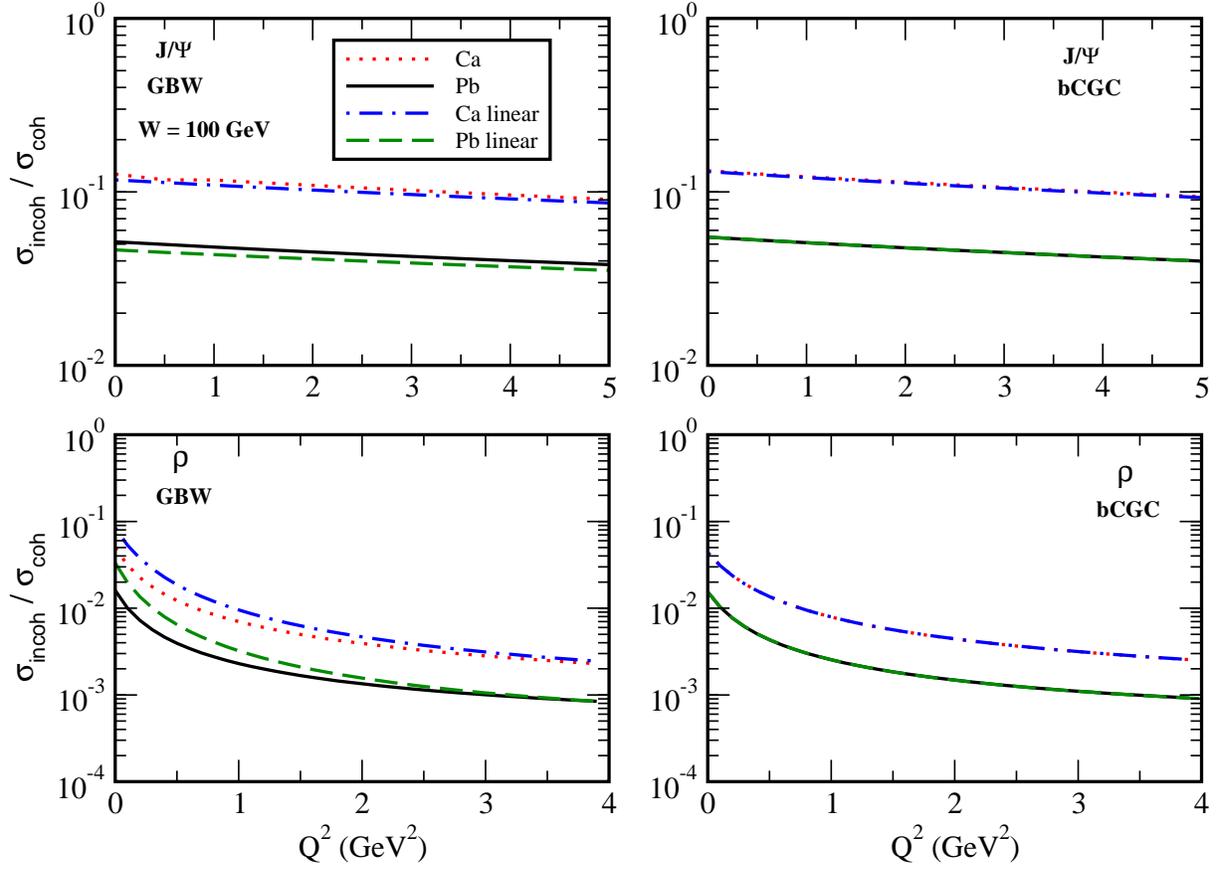} }
\vspace{0.5cm}
\caption{Ratio $\sigma_{incoherent}/ \sigma_{coherent}$ for the nuclear electroproduction of vector mesons as a function of the photon virtuality with the GBW (left panels)  and bCGC (right panels) models at fixed energy ($W$ = 100 GeV).}
\label{fig6}
\end{figure}

Finally, in Figs. \ref{fig5} and \ref{fig6} we present our predictions for the energy and 
$Q^2$ dependences of the ratio between the incoherent and coherent cross sections, 
respectively. The incoherent  contribution is a small fraction of  the coherent 
one and  the ratio decreases with the energy.  Moreover, the ratio is larger for small 
values of the atomic number. These conclusions are weakly model dependent. Concerning the 
$Q^2$ dependence, the ratio decreases at larger $Q^2$, with a stronger dependence in the case  
of the $\rho$ meson, as expected from the discussion above. It is important to emphasize that 
at large $Q^2$ the ratio is larger for $J/\Psi$  than for $\rho$ production. We can conclude 
that although measuring experimentally the intact recoil nucleus (coherent process) in a future 
electron ion collider may be very difficult, our results strongly suggest that this is the 
dominant process in vector meson production, specially when we focus on high energy $e A $ 
reactions. This conclusion agrees with that obtained in \cite{tuc4}.

So far we have been working only  with two phenomenological models for the dipole-proton  scattering amplitude,  GBW and bCGC, and one might wonder
what would be the results obtained with other existing models. In a previous paper we have 
performed a more systematic comparison between different dipole models \cite{kgn05}. Based on 
this comparison, we believe that using other scattering amplitudes would slightly change the   
normalization of our predictions for the cross sections. However all the qualitative features would remain the 
same. In particular the low sensitivity to non-linear effects in the proton, observed in all the curves obtained with the bCGC model,   would  be the same for the other 
models. This can be traced back to the dependence of the saturation scale on the energy and 
dipole pair separation, which determine the speed with which the amplitude goes to one, which is very similar in the modern phenomenological models based on the CGC formalism.






























\section{Conclusions}
\label{sec:conc}

Although a large body of experimental data has been accumulated by the HERA collaborations H1 
and ZEUS over more than one decade in $ep$ collisions, very little is known about vector meson 
production off nuclei at high energies. In this regime we expect a significant contribution 
from saturation physics. In this paper we have estimated the coherent and incoherent cross 
sections for the exclusive vector meson production considering the color dipole approach and 
phenomenological saturation models which describe the scarce $F_2^A$ data as well as the HERA 
data. Our results demonstrate that the coherent production of vector mesons is dominant, with 
a small contribution coming from incoherent processes. Probably in the future $eA$  colliders 
the separation between coherent and incoherent processes will  be difficult. However, in view 
of our results it might be worth trying.

\begin{acknowledgments}

This work was  partially financed by the Brazilian funding 
agencies CNPq and FAPESP. The authors are deeply grateful to M. Munhoz, 
A.J.R. da Silva and R.G. Amorim for providing access to their computing facilities. 

\end{acknowledgments}

\hspace{1.0cm}

\end{document}